\documentstyle[preprint,aps]{revtex}
\tightenlines

\newcommand\mass{\Psi}

\begin{document}

\title{Ensemble Inequivalence in Systems with Long-range Interactions}

\author{Francois Leyvraz\thanks{E-mail: leyvraz@fis.unam.mx. Permanent 
address: Centro de Ciencias F\'\i sicas, Av. Universidad s/n, 62251, 
Cuernavaca, Morelos, Mexico} and 
Stefano Ruffo\thanks{E-mail:ruffo@avanzi.de.unifi.it}}

\address{Dipartimento di Energetica ``S. Stecco", 
Universit\'a di Firenze, Via S. Marta, 
3 I-50139, Firenze, Italy, INFM and INFN, Firenze} 

\date{\today}
\maketitle

\begin{abstract}
Ensemble inequivalence has been observed in several systems. 
In particular it has been recently shown 
that negative specific heat can arise in the microcanonical 
ensemble in the thermodynamic limit for systems with 
long-range interactions.
We display a connection between such behaviour and a 
mean-field like structure of the partition function. Since
short-range models cannot display this kind of behaviour,
this strongly suggests that such systems are necessarily non-mean
field in the sense indicated here. 
We illustrate our results showing an application to the
Blume-Emery-Griffiths model.
We further show that a broad
class of systems with non-integrable interactions are 
indeed of mean-field type in the sense specified, so that
they are expected to display ensemble inequivalence as
well as the peculiar behaviour described above in
the microcanonical ensemble.
\end{abstract}

\vspace{0.3cm}
\noindent PACS number(s): 05.20.Gg, 05.50.+q, 05.70.Fh, 64.60.-i

\section{Introduction}
Particle or spin systems for which the pairwise interaction
potential decays at large distances with a power smaller than
space dimension are called {\it long-range} or {\it non-integrable}.
It has been suggested that at first order phase
transitions such systems should display {\it ensemble inequivalence}
also in the thermodynamic limit~\cite{Thirring,Gross}. A few examples 
where this is explicitly shown, both analytically and numerically, have been 
published~\cite{Thirring,Lynden,Antoni,Barre,Cohen}.
The specific heat, which is always positive in the canonical
ensemble, may become negative in the microcanonical ensemble, and even
temperature jumps when continuously varying the energy may
appear. Negative specific heat was first observed in gravitational
systems during the process of {\it gravothermal collapse}, but
here the situation is made more complex by the singularity of
the interaction at short distances~(for a review, see \cite{Padmanabhan}).
In this paper we present a formal approach within which we can
easily prove ensemble inequivalence, and all the unexpected 
features of the microcanonical ensemble naturally arise. The approach 
is based on the assumption, which we justify afterwards, that a sort 
of Landau free energy can be always introduced for long-range systems
and that it is endowed of good analyticity properties.
A preliminary version of these results has been shortly presented
in a workshop proceedings~\cite{Leyvraz}.

\section{Mean-field and Ensemble Inequivalence}
In this section, we show the main result of this paper. We say that
a system is of {\it mean-field type\/} if it satisfies the following
condition:
\begin{equation}
Z_C(\beta)=\int_{-\infty}^\infty \exp\left[-N\mass(\beta,m)\right]dm,
\label{eq:mean-field}
\end{equation} 
where $Z_C(\beta)$ is the canonical partition function and $\mass(\beta,m)$
is a real analytic function of $\beta$ and $m$. Note that we require this
for finite values of $N$, but that we additionally require
the function $\mass(\beta,m)$ to be independent of $N$. A weaker 
requirement might be that $\mass$ is, in some sense, 
well-behaved in the $N\to\infty$ limit as
far as analyticity is concerned.
However, we shall not pursue this line any further here. 
Note in particular that infinite range models always have this 
property, as follows from their solution using, for example,
the Hubbard--Stratonovich transformation.

From this, it follows almost immediately that ensemble inequivalence 
will in general occur. Indeed, from (\ref{eq:mean-field}) one obtains
the following expression for the phase space volume of the energy
shell, which may be viewed as a kind of microcanonical partition 
function:
\begin{equation}
Z_M(\epsilon)=\int_{-\infty}^\infty\frac{d\lambda}{2\pi}
e^{i\lambda N\epsilon}Z_C(i\lambda),
\label{eq:basic-micro}
\end{equation} 
where $\epsilon$ is the energy per particle. Using analyticity
to rotate the integration contour,
this can be recast in the following form, after inverting the
order of integration: 
\begin{equation}
Z_M(\epsilon)=\int_{-\infty}^\infty dm\int_{-i\infty}^{i\infty}
\frac{d\lambda}{2\pi i}\exp\left[
N(\lambda\epsilon-\mass(\lambda,m)
\right].
\label{eq:double-integral}
\end{equation} 
One can now perform a saddle point integral to estimate the value
of the $\lambda$ integral for large values of $N$. We argue that the
dominant saddle point must lie on the real axis: If it were otherwise, 
we would, because of the reality properties of $\mass$, have two
complex conjugate dominant saddle points, and hence an oscillatory
behaviour of the partition function. However, if the dominant behaviour
of the partition function is oscillatory, it will generally take on 
at least some negative values, which is absurd. We may therefore limit 
ourselves to the consideration of real saddle points. Since the 
integration path is perpendicular to the real axis, however, the minima
along the real $\lambda$ axis of the 
argument of the exponential will correspond to maxima when traversed along 
the $\lambda$ imaginary axis. We are therefore led to the following 
expression for the microcanonical partition function:
\begin{eqnarray}
Z_M(\epsilon)&=&\int_{-\infty}^\infty\exp\left[
N\min_\lambda(\lambda\epsilon-\mass(\lambda,m))
\right]\nonumber\\
&=&\exp\left[
N\max_m\min_\lambda(\lambda\epsilon-\mass(\lambda,m))
\right].
\label{eq:final-micro}
\end{eqnarray} 
From this one immediately obtains for the entropy per particle
\begin{equation}
S_M(\epsilon)=\max_m\min_\lambda[\lambda\epsilon-\mass(\lambda,m))].
\label{eq:entropy-micro}
\end{equation} 
This result can now be compared
with the standard result for the canonical ensemble, which is
obtained as follows: The free energy per particle is found, 
after a straightforward estimate of (\ref{eq:mean-field}) 
via Laplace's method, to be given by
\begin{equation}
F(\beta)=\beta^{-1}\min_m\mass(\beta,m).
\label{eq:free-energy}
\end{equation} 
It is now readily seen, through standard thermodynamic identities,
that the entropy is the Legendre transform of $\beta F(\beta)$
with respect to $\beta$. This then leads to the following 
expression for $S_C(\epsilon)$
\begin{equation}
S_C(\epsilon)=\min_\lambda\max_m[\lambda\epsilon-\mass(\lambda,m))],
\label{eq:entropy-canonical}
\end{equation} 
From these results, a few consequences are immediate:
\begin{enumerate}

\item The two forms of the entropy need not be equal. Indeed, if 
we could prove that $\mass(\lambda,m)$ has a unique extremum under
certain conditions, we could argue for equality. However, we 
know that this is not generally the case: Whenever a phase transition 
occurs, the function $\mass(\lambda,m)$ may have multiple
extrema at least as a function of $m$, thus precluding any simple
statements about the identity of the two entropies.

\item
Quite generally, however, one may say that
\begin{equation}
S_M(\epsilon)\leq S_C(\epsilon).
\label{eq:inequality}
\end{equation} 
This is a general property of such min-max combinations and is trivial
to prove (see Appendix A). It is, however, satisfying, since it suggests
that in some sense the microcanonical entropy is a restricted entropy,
and that the real reason why the two are different is that 
the microcanonical system cannot relax in certain ways, reaching
the maximal possible entropy in the canonical ensemble. 

\item
Let us now consider the behaviour of the function 
$\mass(\lambda,m)$ at the equilibrium points 
$(\lambda_c^\star,m_c^\star)$ and 
$(\lambda_m^\star,m_m^\star)$ for the canonical and 
microcanonical problem respectively. For the 
canonical problem we find the following conditions
for the point $(\lambda_c^\star,m_c^\star)$ to be the 
appropriate extremum
\begin{eqnarray}
\frac{\partial^2\mass}{\partial\lambda^2}&<&0\nonumber\\
\frac{\partial^2\mass}{\partial m^2}&>&0.
\label{eq:stability-canonical}
\end{eqnarray} 
These correspond quite naturally to the usual stability conditions 
for the thermodynamic potential $\mass(\lambda,m)$, which is concave
with respect to $\lambda$ and convex with respect to $m$. 
However, if one goes through the same computation for the
microcanonical ensemble, one finds
\begin{eqnarray}
&&\frac{\partial^2\mass}{\partial\lambda^2}<0\nonumber\\ 
&&\frac{\partial^2\mass}{\partial m^2}-\frac{(\partial^2\mass/
\partial m\partial\lambda)^2}{\partial^2\mass/\partial\lambda^2}
>0.
\label{eq:stability-micro}
\end{eqnarray} 
Note in particular how this implies that the second derivative
of $\mass$ with respect to $m$ can take either sign. In fact, one
verifies that the above two conditions have the following geometric
meaning: The first indicates that $\mass$ is a local maximum in the
$\lambda$ direction, and the second amounts to stating that the 
matrix of second derivatives of $\mass$ has negative determinant, that
is, the extremum must be of saddle-point type. These conditions
are clearly weaker than (\ref{eq:stability-canonical}), so that 
more extrema are allowed for the microcanonical ensemble than
for the canonical.

\item
If we consider the failure of concavity of the entropy as a 
function of $\epsilon$, related to the findings concerning negative
specific heats in certain systems, we again find that this 
can happen in the microcanonical case. Indeed, one sees 
immediately that this cannot occur in the canonical ensemble, 
since there the entropy is given as a minimum of linear functions, 
which is of necessity concave. On the other hand, the microcanonical
expression yields a maximum over concave functions, which need not
be concave. More specifically, an explicit evaluation
of the second derivative of $S_M(\epsilon)$ yields
\begin{equation}
\frac{d^2S_M}{d\epsilon^2}=\frac{\partial^2\mass}{\partial\lambda^2}
-\frac{
(\partial^2\mass/\partial m\partial\lambda)^2
}{
\partial^2\mass/\partial m^2
}.
\label{eq:entropy-2-derivative}
\end{equation} 
Again, since $\partial^2\mass/\partial m^2$ can take both signs,
one sees that the specific heat can do so as well. In fact, combining
this result with (\ref{eq:stability-micro}), one finds that 
the sign of $d^2S_M/d\epsilon^2$ is the opposite of that
of $\partial^2\mass/\partial m^2$. This means that the specific heat
is negative exactly when the value of $m$ corresponding
to microcanonical equilibrium is unstable from the point
of view of the canonical ensemble. 

\item
It has also been found that the temperature $T$ can have a discontinuous 
dependence on $\epsilon$ in the microcanonical ensemble. Again this
readily follows from our formalism: In the canonical ensemble,
one effects a Legendre transformation of the following function
\begin{equation}
F(\lambda)=\max_m\mass(\lambda,m).
\label{eq:def-F}
\end{equation} 
Now $F$ can have discontinuities in its derivative, but it follows from
the analyticity of $\mass$ that it cannot have any straight line segments.
From this follows through well-known considerations that the entropy
can have straight line segments (corresponding to ordinary first-order
phase transitions), but no jumps in its derivative. On the other
hand, the microcanonical entropy $S_M(\epsilon)$ arises as the 
maximum over a set of concave functions, none of which can have
jumps in the derivative. This clearly cannot exclude such
jumps in $S_M(\epsilon)$, which have indeed been found in
specific models.

\item
Finally, we may note that, since the equivalence between ensembles 
is rigorously proved for interactions which are of sufficiently
short-range~\cite{Ruelle}, this argument strongly suggests 
that such systems are by
necessity non mean-field, that is, that such functions as $\mass$
do not exist for such systems, or, more precisely, do not have
the required analyticity properties. The argument is not fully 
rigorous: As pointed out above, it is in principle possible to 
have such an analytic function and yet not to have any cases
in which the two ensembles differ. This would in particular be 
the case if, for all physical parameter values, the function $\mass$ 
had only one extremum. However, it is certainly very suggestive.
\end{enumerate}
\section{Application to the Blume--Emery--Griffiths Model}
Let us show how the above approach works in the case of the 
Blume--Emery--Griffiths (BEG) model~\cite{BEG}, which is a spin Hamiltonian
on a lattice defined as follows:
\begin{equation}
H=\Delta\sum_{i=1}^Ns_i^2
-\frac{J}{2N}\left(\sum_{i=1}^N s_i 
\right)^2,
\label{eq:def-BEG}
\end{equation} 
where the $s_i$ are spin variables taking the values $0, \pm1$. 
$\Delta$ and $J$ are positive coupling constants. This model has
a line of second order phase transitions, which terminates
in a tricritical point and is followed by a line of first-order
phase transitions. This model, since it is in the mean-field limit, 
can be solved exactly using the Hubbard--Stratonovich transformation,
to yield a representation of the partition function of the
form (\ref{eq:mean-field}). One finds
\begin{equation}
\mass(\beta,m)=\frac{J\beta m^2}{2}-\ln\left(
1+2e^{-\beta\Delta}\cosh(\beta Jm)
\right).
\label{eq:BEG-partition}
\end{equation} 
We now show that finding the extremum of 
$\lambda\epsilon-\mass(\lambda,m)$ with respect to $\lambda$
and $m$ is equivalent to looking for the extrema of the entropy
at given energy $\epsilon$ in the microcanonical ensemble.
Indeed, if one defines
\begin{eqnarray}
m&=&\frac{1}{N}\sum_{i=1}^N s_i\nonumber\\
q&=&\frac{1}{N}\sum_{i=1}^Ns_i^2,
\label{eq:def-mq}
\end{eqnarray} 
one readily finds that the entropy and the energy of 
(\ref{eq:def-BEG}) are given by
\begin{eqnarray}
S_M(\epsilon)&=&(1-q)\ln(1-q)+\frac{q+m}{2}\ln\frac{q+m}{2}
+\frac{q-m}{2}\ln\frac{q-m}{2}\nonumber\\
\epsilon&=&\Delta q-\frac{Jm^2}{2}.
\label{eq:def-Se}
\end{eqnarray} 
If one now maximizes
\begin{equation}
S_M(\epsilon)+\mu\left(
\epsilon-\Delta q+\frac{Jm^2}{2}
\right),
\label{eq:lagrange}
\end{equation} 
one finds that this leads to the same values of $\mu$ and $m$ as
the maximization of $\lambda\epsilon-\mass(\lambda,m)$ gives
for $\lambda$ and $m$ respectively. Further, the resulting 
values for the entropy are shown to be equal. These somewhat 
messy computations are performed in Appendix B. It has been recently shown
that in this model all the peculiarities mentioned in Section 2
actually occur~\cite{Barre}.
\section{General Behaviour at Second order Phase Transitions 
and Tricritical Points}
In the preceding Section, we have shown how, in the case of 
a specific model, our formalism was able to re-derive the usual 
microcanonical expressions. However, as these expressions are quite 
cumbersome, it is not possible to derive in a transparent
and explicit way the exact nature of the various thermodynamic
anomalies encountered there. In this Section, we shall pursue
a different approach: We shall consider quite general systems,
but limit ourselves to the region in the vicinity of a phase transition.
Under these circumstances, one can perform a Taylor expansion 
of the function $\mass$ as a low-order polynomial. This then 
allows to carry out a complete study of the stability conditions.

To apply our formalism as developed in Section 1, we require
an approximate expression for the function
\begin{equation}
G(\lambda,m;\epsilon,\Delta)=\lambda\epsilon-\mass(\lambda,m;\Delta),
\label{eq:def-G}
\end{equation} 
where $\Delta$ represents one or more additional
parameters of the Hamiltonian. Since we are interested in 
the vicinity of phase transitions, we are in fact 
concentrating on those regions in which new stationary points
of $G$ appear or disappear. Such regions are found in the vicinity
of the solutions of the following equation:
condition
\begin{equation}
\det D^2G(\lambda,m;\epsilon,\Delta)=0,
\label{eq:hessian}
\end{equation} 
where $D^2G$ means the matrix of second derivatives of $G$ taken
with respect to $\lambda$ and $m$, and the variables $\epsilon$
and $\Delta$ are chosen so that $\lambda$ and $m$ are the corresponding
stationary values. 

Let us first look at the simplest case in which there are no additional 
parameters. If we further assume that the function $G$ is symmetric 
in $m$ with respect to change of sign, one obtains the following result:
it is generically possible for two new minima to arise whenever the second 
derivative of $G$ with respect to $m$ vanishes at the origin. 
If we consider $G$ near this point and measure all quantities with
respect to the critical values, which we therefore set equal to
zero, we obtain the following expression for $G$ in the vicinity
of the transition:
\begin{equation}
G(\lambda,m;\epsilon)=\frac{c_0\lambda^2}{2}+a_2\lambda m^2-m^4+
\lambda\epsilon,
\label{eq:G-taylor}
\end{equation} 
where we have introduced an arbitrary numerical factor to make the coefficient
of the $m^4$ term equal to unity and the minus sign is required by 
the thermodynamic stability of the system. The other constants are
typically of order one. New phenomena may indeed appear when these are
of the same order as $m$, in which case, however, new parameters
$\Delta$ should be introduced and the system should be embedded in one
that has an appropriate multicritical point. 

One now needs to test for the microcanonical stability conditions
(\ref{eq:stability-micro}). The first one, which is also 
necessary for canonical stability, is equivalent to the 
positivity of $c_0$. The second one is written in terms of $G$ as
\begin{equation}
\det D^2G=\det
\left(
\matrix{
&c_0&2a_2m\cr
&2a_2m &2a_2\lambda-12m^2
}
\right)<0,
\label{eq:stability-1}
\end{equation} 
and the equations expressing the stationarity of $G$ with respect
to variations of $\lambda$ and $m$ are given by
\begin{eqnarray}
2a_2\lambda m-4m^3&=&0
\label{eq:stationary-1}\\
c_0\lambda+a_2m^2+\epsilon&=&0.
\label{eq:stationary-2}
\end{eqnarray} 
The equation (\ref{eq:stationary-1}) determines $\lambda$ once $m$ is known,
unless $m$ is zero (paramagnetic phase), in which case 
(\ref{eq:stationary-1}) is a tautology. In this latter case, the condition
(\ref{eq:stability-1}) is equivalent to $a_2\lambda<0$, since $c_0$
is always positive. However, this is just the condition
that ensures the absence of non-zero real solutions
of (\ref{eq:stationary-1}, so that we may say in summary that 
the paramagnetic phase is microcanonically stable 
only if there is no ferromagnetic solution. 

On the other hand, if $m\neq0$, it follows that 
\begin{equation}
a_2\lambda=2m^2,
\label{eq:trivial-1}
\end{equation} 
from which it is immediate that (\ref{eq:stability-1}) always holds good.
Therefore, the ferromagnetic solution is always stable whenever it exists. 
For this reason, there is no possibility of
ensemble inequivalence near a generic second-order phase transition.

The issue of tricritical points is a bit more subtle, since in this case 
we may have indeed ensemble inequivalence. The simplest case, 
to which we limit ourselves, is that in which a single additional parameter
$\Delta$ is introduced and one considers the case where the second
and the fourth derivative of $G$ vanish simultaneously
when $m=0$. We assume that this occurs at $\epsilon=\Delta=0$, 
and we choose the value of $\lambda$ which corresponds to 
the tricritical point to be zero as well. 
Again we assume $G$ symmetric in $m$ around zero. In this case 
the Taylor expansion is given by
\begin{equation}
G(\lambda,m;\epsilon,\Delta)=\frac{c_0\lambda^2}{2}
+(a_2\lambda+b_2\Delta)m^2+(a_4\lambda+b_4\Delta)m^4
-m^6+\lambda\epsilon.
\label{eq:Taylor-2}
\end{equation} 
The equations determining $\lambda$ and $m$ are now
\begin{eqnarray}
(a_2\lambda+b_2\Delta)+2(a_4\lambda+b_4\Delta)m^2-3m^4&=&0
\label{eq:stationary-3}\\
c_0\lambda+a_2m^2+a_4m^4+\epsilon&=&0.
\label{eq:stationary-4}
\end{eqnarray} 
Here (\ref{eq:stationary-3}) holds only if $m\neq0$, otherwise it
should be discarded. From (\ref{eq:stationary-4}) one obtains
\begin{equation}
\lambda=-\frac{a_2m^2+\epsilon}{c_0}+O(m^4).
\label{eq:m-lambda}
\end{equation}
If one substitutes (\ref{eq:m-lambda}) into (\ref{eq:Taylor-2}), one
obtains a term of order $m^4$ with prefactor $-a_2^2/c_0$, which 
is of order one and is not compensated by any other term of similar
order of magnitude. Thus we see that in the microcanonical ensemble
the point corresponding to the canonical tricritical point is not
tricritical any more. Rather, since the $m^4$ term has negative 
sign, we do not need the $m^6$ term for stabilization any more 
and we are led back to the case of second order transitions
discussed above. From this fact it immediately follows that 
ensemble equivalence cannot hold near canonical tricritical 
points. On the other hand, there may be {\it microcanonical\/}
tricritical points, in which various microcanonical equilibria merge
in the same way as in a canonical tricritical point. The study of
such points is reserved to a later study. 
\section{Models with Long-range Interactions}
In this Section, we show that a large class of models having non-integrable
interactions, actually satisfy our criterion (\ref{eq:mean-field})
and can hence be expected to display the whole gamut of
phenomena discussed in Section 2. For definiteness' sake we restrict
ourselves to spin models on a lattice. 
Extensions to more general cases is presumably unproblematic.
Consider the following Hamiltonian
\begin{equation}
H\left[s(\vec i)\right]=L^{-(d-\alpha)}\sum_{\vec k, \vec l}
\frac{s(\vec k)s(\vec l)}{|\vec k-\vec l|^\alpha}+
\sum_{\vec k}V\left[s(\vec k)\right].
\label{eq:hamiltonian}
\end{equation} 
Here the $s$ are spins which run over a discrete set $S$, the indices $\vec k$
run over a $d$-dimensional lattice and $\alpha$ is an exponent between zero
and $d$. The normalization of the interaction by $L^{-(d-\alpha)}$
guarantees that the Hamiltonian is in fact {\it extensive}.
Note that the BEG model treated above, see (\ref{eq:def-BEG}), 
corresponds to a special case of (\ref{eq:hamiltonian}),
in which $\alpha=0$. 

To evaluate the partition function, we divide the volume in a 
large number of cells, such that the following conditions are
satisfied: first, the cells are large enough that a partition
function involving only the one-body terms can be evaluated accurately 
using saddle-point techniques; second, the cell should be small enough 
with respect to the whole sample, so that the interaction 
between the spins of one cell is negligible compared to the interaction 
with all other cells. Due to the non-integrable nature of 
the interaction, this can always be achieved by making the 
sample sufficiently large. We denote by $\vec x, \vec y$ the centers
of the cells and introduce the coarse-grained variables:
\begin{equation}
\rho(\vec x)=\frac{1}{v}\sum_{\vec k\in C(\vec x)}s(\vec k).
\label{eq:def-rho}
\end{equation} 
Here $C(\vec x)$ denotes the cell at $\vec x$ and $v$ its volume. 
We further define $\mass(\beta,\rho)$ for a system
of volume $v$ by
\begin{equation}
\exp[-N\mass(\beta,\rho)]=\sum_{s(\vec k)}
\delta\left[\frac{1}{v}\sum_{\vec k}s(\vec k)-\rho\right]
\exp\left[-\beta\sum_{\vec k}V\left(s(\vec k)\right)\right].
\label{eq:def-mass}
\end{equation} 
In order to express the partition function in terms of the 
coarse-grained variables, we need an expression for the interaction
energy in terms of $\rho(\vec x)$. This is obtained by the following 
consideration: As a consequence of the normalization, the interaction
between any spin and those of its own cell or of nearby cells
can be neglected. Since one only needs to consider distant spins, 
it is sufficient to use the first term of a Taylor expansion
for the energy, which leads to the expression
\begin{equation}
E=L^{-(d-\alpha)}\sum_{\vec x,\vec y}\frac{\rho(\vec x)\rho(\vec y)}{
|\vec x-\vec y|^\alpha}.
\label{eq:energy}
\end{equation} 
From this one obtains the following expression for the 
partition function
\begin{equation}
\sum_{\rho(\vec x)}\exp\left[
-\frac{\beta}{L^{d-\alpha}}
\sum_{\vec x,\vec y}\frac{\rho(\vec x)\rho(\vec y)}{
|\vec x-\vec y|^\alpha}-N\sum_{\vec x}\mass(\beta,\rho(\vec x))
\right].
\label{eq:partition}
\end{equation} 
Here the sum extends over all functions $\rho(\vec x)$, which have 
integral $N$ and $N$ is the total number of cells. This can now be
evaluated straightforwardly using Laplace's method. The main 
contribution comes therefore from solutions of the equation
\begin{equation}
\frac{\partial\mass}{\partial\rho}
(\beta,\rho(\vec x))=-\frac{\beta}{L^{d-\alpha}}
\sum_{\vec y}\frac{\rho(\vec y)}{
|\vec x-\vec y|^\alpha
}.
\label{eq:extremum}
\end{equation} 
We now show that the right-hand side is independent of $\vec x$. From this
follows that the dominant contributions come from constant
density profiles. One can therefore replace the functional integration
in (\ref{eq:partition}) by an ordinary integration over
the constant value of $\rho(\vec x)$ and one has cast the partition
function in the form (\ref{eq:mean-field}).

To show the constancy of the right-hand side of (\ref{eq:extremum}),
it is enough to note the following: The difference between the 
right-hand side evaluated at $\vec x$ and evaluated at $\vec x+\vec a$
contains two types of terms: First, contributions due to cells 
near $\vec x$ or $\vec x+\vec a$. These give a finite number of 
contributions of order one to the sum, and no contribution
to the right-hand side itself, because of the normalization.
Secondly, one has distant terms, the contribution of which is estimated
by a Taylor expansion. However, it is clear that in leading order, these terms
are identical for $\vec x$ and $\vec x +\vec a$, thus showing the result.
\section{Conclusions and perspectives}
Having shown that the mean-field formalism analysed in Section 2 already
contains all the peculiarities of ensemble inequivalence, a first future
direction of research might be to obtain exact, approximate or even numerical
solutions of well chosen models with slowly decaying interactions, of
the type discussed in Section 5. The aim will be to test the behavior
of physical quantities, like the specific heat, as the exponent of
the decaying interaction is varied from long-range to short-range
coupling, similarly to what has been done for coupled rotators 
models~\cite{Anteneodo,Giansanti}. A related issue is the study of
the behaviour of finite systems as a function of the ratio between
the system size and the range of interaction.
A further direction of research has to do with non-equilibrium properties.
For instance, one might ask what happens when two
systems of comparable size, both having negative specific heat, 
are put in contact with each other. This removes the
energy conservation constraint of the two systems. What is the
behavior of the full system in this case? We might expect 
the microcanonical entropy to come closer to the
canonical, although remaining bounded from above by it.
This would yield the paradoxical result that the coupling
of two systems with identical intensive parameters leads to
an irreversible increase of entropy. 

Finally, we have not at all mentioned in this paper non-equilibrium 
dynamical properties of long-range systems, which might well be analysed
in the context of the thermodynamic formalism here introduced.
For instance, the coherent clustering behavior observed in the
antiferromagnetic mean-field XY model~\cite{Dauxois}, and recently
found to be a non-equilibrium phenomenon~\cite{Firpo}, might 
well find some explanation in the context of the current formalism. 
\section*{Acknowledgements}
We thank E.G.D.~Cohen, J.~Barr\'e, T.~Dauxois, M.-C.~Firpo, 
D.H.E.~Gross, S.~Lepri, A.~Torcini and E.~Votyakov for stimulating 
discussions. This work is supported in part by the 
MURST-COFIN00 project {\it Chaos and localization in 
classical and quantum mechanics}, by INFN, the University
of Florence as well as by CONACyT project 32173-E and 
DGAPA IN112200.
\appendix
\section{}
In this appendix, we show quite generally the inequality
\begin{equation}
\max_x\min_y f(x,y)\leq\min_y\max_x f(x,y),
\label{eq:inequality-general}
\end{equation} 
from which (\ref{eq:inequality}) follows immediately. Assume that 
the two extrema are attained at the two points $(x_1,y_1)$ and
$(x_2,y_2)$ respectively, that is
\begin{eqnarray}
f(x_1,y_1)&\leq& f(x_1,y)\nonumber\\
f(x_2,y_2)&\geq& f(x,y_2)
\label{eq:extrema-def}
\end{eqnarray} 
From this follows
\begin{equation}
f(x_1,y_1)\leq f(x_1,y_2)\leq f(x_2,y_2),
\label{eq:trivial}
\end{equation} 
which implies the inequality (\ref{eq:inequality-general}).
\section{}
Here we complete the proof of the equivalence between the 
two ways of computing the microcanonical entropy of the BEG model, by 
performing the missing computations. The equations arising from the
minimization of (\ref{eq:BEG-partition}) are
\begin{eqnarray}
\epsilon-\frac{Jm^2}{2}+2\frac{
-\Delta e^{-\Delta\lambda}\cosh Jm\lambda+Jme^{-\Delta\lambda}
\sinh Jm\lambda
}
{
1+2e^{-\Delta\lambda}\cosh Jm\lambda
}&=&0
\nonumber\\
-m+2\frac{
e^{-\Delta\lambda}\sinh Jm\lambda
}
{
1+2e^{-\Delta\lambda}\cosh Jm\lambda
}&=&0,
\label{eq:A1}
\end{eqnarray} 
which are readily combined to yield
\begin{eqnarray}
&&m=2\frac{
e^{-\Delta\lambda}\sinh Jm\lambda
}
{
1+2e^{-\Delta\lambda}\cosh Jm\lambda
}\nonumber\\
&&\epsilon+\frac{Jm^2}{2}-\Delta e^{-\Delta\lambda}m\coth Jm\lambda
=0.
\label{eq:A2}
\end{eqnarray} 
On the other hand, finding the extrema of (\ref{eq:lagrange}) leads to the 
following conditions
\begin{eqnarray}
\ln(1-q)+\ln\frac{q+m}{2}+\ln\frac{q-m}{2}-\mu\Delta&=&0
\label{eq:A3}\\
\ln\frac{q+m}{2}-\ln\frac{q-m}{2}+2J\mu m&=&0
\label{eq:A4}\\
\epsilon+\frac{Jm^2}{2}-\Delta q&=&0.
\label{eq:A5}
\end{eqnarray} 
From (\ref{eq:A4}) one obtains
\begin{equation}
q=m\coth Jm\mu
\label{eq:A6}
\end{equation} 
If one now substitutes (\ref{eq:A6}) into (\ref{eq:A3}) and
(\ref{eq:A5}), one obtains the equations (\ref{eq:A2}), where $\lambda$
must be identified with $\mu$. We therefore see that the two sets 
of equations are fully equivalent. The verification that the
value of the entropy is the same in both cases is now straightforward
algebra.


\begin{thebibliography}{99}
\bibitem{Thirring} P.~Hertel and W.~Thirring, {\em Ann. of Phys.} 
{\bf 63} (1971) 520. 
\bibitem{Gross}D.H.E.~Gross and E. Votyakov, 
{\em Eur. Phys. J. B} {\bf 15} (2000) 115.
\bibitem{Lynden}R.M. Lynden-Bell,
{\em Mol. Phys.} {\bf 86} (1995) 1353.
\bibitem{Antoni} A.~Torcini and M.~Antoni, 
{\em Phys. Rev. E} {\bf 59} (1999) 2746.
\bibitem{Barre} J. Barr\'e, D. Mukamel and S. Ruffo, 
{\em Phys. Rev. Lett.} {\bf 87} (2001) 030601.
\bibitem{Cohen} I. Ispolatov and E.G.D. Cohen, 
{\em Physica A} {\bf 295} (2001) 475.
\bibitem{Padmanabhan} T.~Padmanabhan, {\em Phys. Rep.} {\bf 188} (1990) 285.
\bibitem{Leyvraz} F. Leyvraz and S. Ruffo, preprint (2001).
\bibitem{Ruelle} D. Ruelle, {\em Helv. Phys. Acta} {\bf 36} (1963) 183 
\bibitem{BEG} M. Blume, V.J. Emery and R.B. Griffiths, {\em Phys. Rev. A}
{\bf 4} (1971) 1071.
\bibitem{Anteneodo} F.~Tamarit and C.~Anteneodo, {\em Phys. Rev. Lett.} 
{\bf 84} (2000) 208;  C. Anteneodo and C. Tsallis, {\em Phys. Rev. Lett} 
{\bf 80} (1998) 5313.
\bibitem{Giansanti} A. Campa, A.~Giansanti and D. Moroni, 
{\em Phys. Rev. E} {\bf 62} (2000) 303.
\bibitem{Dauxois} J. Barr\'e, T. Dauxois and S. Ruffo, {\em Physica A}
{\bf 295} 254 (2001); T. Dauxois, P. Holdsworth and S. Ruffo, 
{\em Eur. Phys. J B} {\bf 16} (2000) 659.
\bibitem{Firpo} F. Leyvraz, M.-C. Firpo and S. Ruffo, preprint (2001). 
\end{thebibliography}
\end{document}